\documentclass[12pt]{article}
\advance\textwidth by +1.0in
\advance\textheight by +1.0in
\advance\oddsidemargin by -0.5in
\advance\evensidemargin by -1.0in
\advance\topmargin by -0.5in
\def\ov#1{\overline{#1}}

\def\wt#1{\widetilde{#1}}
\def\vb#1{\mbox{\boldmath$#1$}}
\def\pd#1#2{\frac{\partial #1}{\partial #2}}

\def\wh#1{\widehat{#1}}
\def\bdot{\,\vb{\cdot}\,}
\def\btimes{\,\vb{\times}\,}

\newcommand{\bc}{\begin{center}}
\newcommand{\ec}{\end{center}}
\newcommand{\bt}{\begin{tabbing}}
\newcommand{\et}{\end{tabbing}} 
\newcommand{\be}{\begin{eqnarray*}}
\newcommand{\ee}{\end{eqnarray*}}

\newcommand{\no}{\noindent}

\begin{document}

\begin{flushright}
January 30, 2003
\end{flushright}

\bc
{\Large {\bf Mini-conference on Hamiltonian and Lagrangian methods in fluid and plasma physics}} \\

\vspace*{0.5in}

Alain J.~Brizard \\
{\it Saint Michael's College, Colchester, Vermont 05439}

and

Eugene R.~Tracy \\
{\it College of William and Mary, Williamsburg, Virginia 23187-8795}
\ec

A mini-conference on Hamiltonian and Lagrangian methods in fluid and plasma physics was held on November 14, 2002, as part of the 44$^{th}$ meeting of the Division of Plasma Physics of the American Physical Society.  This paper summarizes the material presented during the talks scheduled during the Mini-Conference, which was held to honor Allan Kaufman on the occasion of his 75$^{th}$ birthday.

\vfill

\no

\vfill\eject

\no
{\sf I. INTRODUCTION}

\vspace*{0.2in}

The languages of Hamiltonian and Lagrangian mechanics permeate much of fluid and plasma physics. On the one hand, Hamiltonian methods provide powerful analytical and numerical tools for the investigation of the motion of charged 
particles in complex electromagnetic fields and the linear and nonlinear stability of plasma and fluid equilibria. On the other hand, Lagrangian (or variational) formulations of fluid and plasma physics provide clear guiding principles for the systematic dynamical reduction of fluid and plasma dynamics through the asymptotic elimination of fast degrees of freedom and the construction of adiabatic invariants either by Lagrangian averaging or by Lie-transform methods. Moreover, variational formulations facilitate the derivation of important conservation laws (e.g., energy-momentum, angular momentum, and wave action) through the Noether method. 

The present paper summarizes the oral talks presented at the Mini-Conference on Hamiltonian and Lagrangian methods in fluid and plasma physics held at the 2002 annual meeting of the Division of Plasma Physics of the American Physical Society. The paper is organized by topic and not the chronological order in which the talks were given. The organization is, therefore, as follows: Hamiltonian methods (Sec.~II), Lie-transform methods (Sec.~III), Lagrangian methods (Sec.~IV), Lagrangian chaos (sec.~V), and a Summary of the Mini-Conference (Sec.~VI). Because this is a summary of a topical mini-conference, and not a review paper, no attempt has been made to provide a comprehensive list of references. Instead, references to the relevant review papers have been provided, and citations to original sources can be found there.

\vspace*{0.2in}

\no
{\sf II. HAMILTONIAN METHODS}

\vspace*{0.2in}

Applications of Hamiltonian methods relie on the existence of a Hamiltonian function (or functional) $H(\vb{\zeta},t)$, which depends on a set of coordinates (or fields) $\zeta^{k}\; (k = 1,...,N)$, and a Poisson bracket structure $\{\;,\;\}$ such that Hamilton's equations are expressed as $d\zeta^{k}/dt = \{\zeta^{k},\; H(\vb{\zeta},t)\}$. Here, the Poisson bracket must satisfy (i) the antisymmetry property $\{ f,\; g\} = -\,\{ g,\; f\}$, (ii) the Leibnitz property $\{ (f\,g),\; h \} = f\;\{ g,\; h\} + \{f,\; h\}\,g$, and (iii) the Jacobi condition $\{f,\; \{g,\; h\}\} = -\,\{g,\; \{h,\; f\}\} - \{h,\; \{f,\; g\}\}$, where $(f, g, h)$ are arbitrary functions (or functionals). Whether the coordinates $\vb{\zeta}$ are canonical or not determines additional features of the Hamiltonian dynamics. Although the Poisson bracket structure of a canonical Hamiltonian formulation has a simple form, canonical coordinates suffer from the problem of being unphysical in most cases (e.g., they are gauge dependent). The exploitation of noncanonical Hamiltonian formulations was pioneered by the work of Littlejohn on the Hamiltonian theory of guiding-center motion \cite{RGL}. The physical transparency of noncanonical coordinates comes at the cost of the complexity of the Poisson bracket structure, although its derivation is solidly grounded mathematically. An important consequence of the noncanonical Poisson bracket structure is the existence of Casimir invariants $C(\vb{\zeta})$, for which $\{ C(\vb{\zeta}),\; F(\vb{\zeta})\} = 0$ for any function $F(\vb{\zeta})$. For example, the noncanonical Lie-Poisson bracket structure $\{\;,\;\}_{f}$ for the Vlasov equation is an antisymmetric 
bilinear operator on the space of Vlasov distributions $f(z)$ in six-dimensional phase space defined as \cite{VP_Morrison}
\[ \{ F,\; G\}_{f} \;=\; \int\; f(z) \left\{ \frac{\delta F}{\delta f(z)},\; 
\frac{\delta G}{\delta f(z)} \right\}\; d^{6}z, \]
where $F[f]$ and $G[f]$ are two arbitrary functionals and $\{\;,\;\}$ is the Poisson bracket structure on six-dimensional phase space. Casimir invariants for the Vlasov equation 
are of the form $C[f] = \int \chi(f(z))\,d^{6}z$, where $\chi(f)$ is an arbitrary function of $f$ and $\{ C[f],\; F[f]\}_{f} = 0$ for all functionals $F$. Casimir invariants \cite{footnote_1} play an important role in defining relative Hamiltonian equilibria and in investigating linear and nonlinear stability. The present Section focuses on some applications of Hamiltonian methods in fluid and plasma physics.

Phil Morrison, the Mini-Conference keynote speaker, addressed the important question associated with the significance of the existence of a Hamiltonian structure or a variational (Lagrangian) structure for any dissipationless dynamical system \cite{Morrison_RMP}. First, the existence of a variational structure serves as a guiding principle for deriving exact and reduced dynamical equations \cite{Morrison_RMP} and their associated conservation laws. Several variational formulations for fluids and plasmas exist; the variational formulations for the Vlasov equation are reviewed by Ye and Morrison \cite{VP} while the variational formulations for ideal fluids are reviewed by Holm, Marsden, and Ratiu \cite{EP_review}. Next, from 
a practical point of view, the existence of a noncanonical Hamiltonian structure reveals the existence of Casimir invariants for a continuous Hamiltonian system and, thus, allow the investigation of the stability of relative Hamiltonian equilibria \cite{Morrison_RMP}. Through the use of Hamiltonian methods, the stability properties of numerous dissipationless kinetic, hydrodynamic, and magnetohydrodynamic systems have been investigated through the Lyapunov \cite{Holm_stability}, Energy-Casimir \cite{EC}, and Dynamical Accessibility \cite{DA} methods. Another significant application of Hamiltonian and Lagrangian methods in fusion plasma physics lies with the Hamiltonian dynamics of magnetic field lines and the existence of closed magnetic surfaces \cite{magnetic}. 

Next, Morrison reviewed the variational structure of the exact Vlasov equation, beginning with the pioneering work of Low \cite{Low} and culminating with recent works by Cendra {\it et al.} \cite{Cendra_etal} and Brizard \cite{Brizard_VPV}, and the Hamiltonian and Lagrangian formulations of several exact and reduced fluid models (e.g., ideal MHD with flows \cite{Frieman_Rotenberg} and reduced MHD \cite{Morrison_Hazeltine}). Morrison concluded his presentation with a functional-integral derivation of the fluctuation spectrum of a Vlasov-Poisson plasma, again highlighting the usefulness of the Hamiltonian formalism.

Allan Kaufman reviewed Boghosian's work \cite{BMB} on the covariant Lagrangian formulations of guiding-center and oscillation-center plasmas. A covariant variational principle is formulated for Vlasov-Maxwell plasma dynamics in terms of electromagnetic fields/potentials in four-dimensional space-time (with coordinates $r^{\mu}$) and particle orbits in eight-dimensional phase space (with coordinates $z^{\alpha} = (r^{\mu}, p_{\mu})$). The advantages of a covariant formulation include a frame-independent representation in which polarization and magnetization effects are treated in a unified manner. The covariant Hamiltonian and Lagrangian formulations of charged-particle dynamics proceed as follows. First, the covariant Hamiltonian structure of charged particle motion in electromagnetic fields is given in terms of the covariant Hamiltonian $H(z) = p^{\mu}p_{\mu}/2m \equiv - \,mc^{2}/2$ and the covariant Poisson bracket
\begin{equation}
\{ f(z),\; g(z) \} \;=\; \pd{f(z)}{r^{\mu}}\,\pd{g(z)}{p_{\mu}} \;-\;
\pd{f(z)}{p_{\mu}}\,\pd{g(z)}{r^{\mu}}  \;+\; \frac{e}{c}\;\pd{f(z)}{p_{\mu}}\;
F_{\mu\nu}(r)\;\pd{g(z)}{p_{\nu}},
\label{eq:cov_PB}
\end{equation}
where $f(z)$ and $g(z)$ are arbitrary extended-phase-space functions, and $F_{\mu\nu}(r) = \partial_{\mu}A_{\nu}(r) - \partial_{\nu}A_{\mu}(r)$ denotes the electromagnetic field. We note here that the electromagnetic field appears only in 
the Poisson bracket (\ref{eq:cov_PB}), which is derived from the covariant phase-space Lagrangian $\Gamma = (p_{\mu} + e\,A_{\mu}/c)\,dr^{\mu}$, and that the covariant Hamiltonian is actually a Lorentz invariant \cite{Brizard_Chan}. 

The covariant perturbation analysis now proceeds as follows. First, for slowly-varying fields $F_{(0)\mu\nu}$ (with ${\bf E}_{(0)}\bdot{\bf B}_{(0)} = 0$), Lie-transform methods lead to a covariant {\it guiding-center} (gc) description 
($\Gamma \rightarrow \Gamma_{gc}$ and $H \rightarrow H_{gc}$). Next, a small-amplitude eikonal wave 
\[ F_{\mu\nu} \;=\; F_{(0)\mu\nu} +  \epsilon\,\left[\; i \left( k_{\mu}\,\wt{A}_{\nu} - k_{\nu}\,\wt{A}_{\mu}\right)
\,\exp(i\epsilon^{-1}\theta) \;+\; {\rm c.c.} \;\right] \;+\; {\cal O}(\epsilon^{2}) \]
is introduced, where $\epsilon \ll 1$ is the eikonal small parameter while the eikonal amplitude $\wt{A}_{\mu}$, the eikonal phase $\theta$, and the wave four-vector $k_{\mu} = \epsilon^{-1}\partial_{\mu}\theta$ are slowly-varying functions of the space-time guiding-center coordinates, so that the covariant guiding-center phase-space Lagrangian becomes $\Gamma_{gc} \rightarrow \Gamma_{gc} + \epsilon\,\Gamma_{w}$. The wave fast time scale is asymptotically eliminated through a near-identity phase-space transformation from guiding-center coordinates to {\it oscillation-center} (oc) coordinates, expressed as asymptotic series expansions in powers of $\epsilon$ and denoted $Z(z,\epsilon)$. The phase-space transformation also induces a transformation on the Hamiltonian $(H_{gc} \rightarrow K_{oc} = T_{\epsilon}^{-1}H_{gc} = K_{0} + \epsilon^{2}\,K_{2})$, where 
the covariant quadratic ponderomotive Hamiltonian for oscillation-centers is
\begin{equation}
K_{2}(Z;\,{\sf A},{\sf k}) \;=\; \wt{A}_{\mu}^{*}({\sf R})\,K^{\mu\nu}(Z;\,{\sf k})\,
\wt{A}_{\nu}({\sf R}).
\label{eq:pond_Ham}
\end{equation}
Here, the phase-space $4 \times 4$ tensor $K^{\mu\nu}(Z; {\sf k})$ appears in the definition of the linear susceptibility
\[ \chi^{\mu\nu}({\sf x}) \;=\; -\;\int d^{8}Z\,\delta^{4}({\sf R} - {\sf x})\;f_{oc}(Z)\;
K^{\mu\nu}(Z; {\sf k}), \]
where $f_{oc}(Z)$ denotes the oscillation-center Vlasov distribution, which in turn leads to 
the so-called $K-\chi$ theorem \cite{K_chi}
\[ -\;\frac{\delta\chi^{\mu\nu}}{\delta f_{oc}} \;=\; K^{\mu\nu} \;=\;
\frac{\partial^{2}K_{2}}{\partial\wt{A}_{\mu}^{*}\,\partial\wt{A}_{\nu}}, \]
which relates the oscillation-center ponderomotive Hamiltonian to the linear susceptibility.
Lastly, Kaufman emphasized the role played by the Noether method in deriving conservation laws relating the self-consistent evolution of wave, background plasma, and background field \cite{BMB,Kaufman_Holm,Similon}.

Diego del-Castillo-Negrete presented work on the role of self-consistent chaos in mean-field Hamiltonian models of fluids and plasmas. Mean-field Hamiltonian theories are intermediate between test-particle Hamiltonian theories, in which particles interact with wave structures 
but not with each other, and exact Hamiltonian theories, in which self-consistent fields produced by particles are taken into account in wave-particle interactions. Mean-field Hamiltonian models can be used to describe the collective dynamics of marginally stable fluids (e.g., the dynamics of localized vortices with negative and positive circulation in shear flows) and plasmas (e.g., the self-consistent evolution of localized electron holes and clumps in phase space \cite{DCN,A}). During his talk, del-Castillo-Negrete presented the Hamiltonian formulation of the single-wave model (SWM) equations describing the interaction of $N$ particles with phase-space coordinates $\{(x_{j},\; u_{j}); j = 1,..., N\}$ with a single wave (characterized by a time-dependent complex-valued amplitude $a$). Here, the Hamiltonian for the particles is
\begin{equation}
H = \sum_{k = 1}^{N} \left[\; \frac{u_{k}^{2}}{2} \;-\; a(t)\;\exp(i\,x_{k}) \;-\; a^{*}(t)\;\exp(-i\,x_{k}) \;\right],
\label{eq:single_wave}
\end{equation}
so that Hamilton's equations $dx_{j}/dt = \partial H/\partial u_{j}$ and $du_{j}/dt = -\,\partial H/\partial x_{j}$ 
describe the particle dynamics, while the wave dynamics is represented by the equation
\begin{equation}
\sigma\;\frac{da}{dt} \;-\; i\,U\;a \;=\; \frac{i}{N}\; \sum_{k}\; \Gamma_{k}\;\exp(-i\,x_{k}),
\label{eq:a_eq}
\end{equation}
where $(\sigma, U, \Gamma_{k})$ are model parameters (e.g., clumps and holes are labeled by $\Gamma_{k} > 0$ and 
$\Gamma_{k} < 0$, respectively). The SWM equations (\ref{eq:single_wave})-(\ref{eq:a_eq}) have their origin in the study of the beam-plasma instability \cite{B}. More recently, the system has been systematically derived under more general conditions as a basic model describing the weakly nonlinear dynamics of marginally-stable Vlasov-Poisson plasmas and localized vorticity perturbations in shear flows \cite{DCN}. Also, the SWM bears interesting similarities with Hamiltonian models used in the study of long-range interacting systems in statistical mechanics \cite{C}.

Numerical simulations based on the single-wave model equations (\ref{eq:single_wave}), with finite and infinite $N$, show the existence of coherent, rotating dipole states. The coherence of the dipole is explained in terms of a parametric resonance betwen the rotation frequency of the macroparticles and the oscillation frequency of the self-consistent mean field \cite{D}. The role of self-consistent chaos in the formation and destruction of phase-space coherent structures was emphasized by del-Castillo-Negrete. Lastly, for some initial conditions, del-Castillo-Negrete showed that the mean-field exhibits a self-consistent elliptic-hyperbolic bifurcation that leads to the destruction of the dipole and violent mixing 
of the phase space (see Ref.~\cite{C} and Section V on Lagrangian Chaos).

\vspace*{0.2in}

\no
{\sf III. LIE-TRANSFORM METHODS}

\vspace*{0.2in}

The Lie-transform methods traditionally used in Hamiltonian perturbation theory involve near-identity (or 
{\it infinitesimal}) canonical transformations. Practical applications of such {\it perturbative} Lie-transform methods normally involve truncation of asymptotic expansions in powers of a small dimensionless parameter $\epsilon$ at finite order and, consequently, involve moderate algebraic complexity. {\it Non-perturbative} Lie-transform methods, which involve 
{\it finite} canonical transformations, are used in Hamiltonian orbit theory (associated with accelerator design, for example) and are, therefore, typically associated with tremendous algebraic complexity (to the point of requiring the use 
of symbolic manipulation algorithms). The present Section focuses on Alex Dragt's presentation on applications of non-perturbative Lie-transform methods to the technological challenges associated with the construction of powerful particle accelerators \cite{Lie_acc1,Lie_acc2}, high-resolution electron microscopes \cite{Lie_microscope}, and light optical devices \cite{Lie_op1,Lie_op2}.

The Lie-algebraic tools used in accelerator physics and geometric optics include symplectic matrices, symplectic maps, and Lie transformations. A matrix ${\sf M}$ is symplectic if it satisfies the condition ${\sf M}^{T}\cdot{\sf J}\cdot{\sf M} = {\sf J}$, where ${\sf J}$ represents the canonical symplectic matrix; a map ${\cal M}$ is symplectic if its Jacobian matrix is symplectic; and a (finite) Lie transformation is defined as
\begin{equation} 
(\exp\mbox{:f:})\,g \;=\; g \;+\; \{f,\; g\} \;+\; \frac{1}{2} \{ f,\;\{f,\; g\}\,\} 
\;+\; \cdots,
\label{eq:Lie}
\end{equation}
for any phase-space functions $f$ and $g$ (using Dragt's notation $\mbox{:f:}g = \{ f,\; g\}$ for Lie operators). 

With these Lie-algebraic tools in hand, we now contemplate their applications. First, any symplectic map ${\cal M}_{f}$ can be factorized as 
\[ {\cal M}_{f} \;=\; (\exp\mbox{:f$_{2}$:})\,(\exp\mbox{:f$_{3}$:})
\,(\exp\mbox{:f$_{4}$:})\,\cdots, \]
where f$_{n}$ is a homogeneous polynomial of degree $n$ in the phase-space coordinates, 
while its inverse map is defined as ${\cal M}_{f}^{-1} = \cdots (\exp-\mbox{:f$_{4}$:})\,
(\exp-\mbox{:f$_{3}$:})\,(\exp-\mbox{:f$_{2}$:})$. Here, factorization is facilitated by the Baker-Campbell-Hausdorff (BCH) theorem 
\begin{equation}
(\exp\mbox{:f:})\;(\exp\mbox{:g:}) \;=\; (\exp\mbox{:h:}), 
\label{eq:BCH}
\end{equation}
which states that a new Lie operator $\mbox{:h:}$ can constructed from the Lie operators $\mbox{:f:}$ and $\mbox{:g:}$ through the BCH formula
\[ h \;=\; f \;+\; g \;+\; \frac{1}{2}\;\{f,\; g\} \;+\; \left. \left. \frac{1}{12} \right(
\{ f,\; \{f,\; g\} \} \;+\; \{g,\; \{g,\; f\} \} \right) \;+\; \cdots \]
Proof of the BCH theorem (\ref{eq:BCH}) relies on the definition (\ref{eq:Lie}) of the Lie transform and the fact that the Poisson bracket $\{\;,\;\}$ satisfies the antisymmetry property and the Jacobi identity.

Next, the Hamiltonian orbits of charged particles moving in an electromagnetic field (or light rays) propagating through an assembly of beam-line elements (or lenses) are generated by symplectic maps (i.e., finite canonical transformations). Since the guiding action of electromagnetic fields has strong similarities with the guiding action of lenses, the Lie-algebraic tools defined above can thus find applications both in accelerator physics and in geometric optics. First, a series of {\it elements} (electromagnetic multipoles or lenses) can be represented by the map product ${\cal M} = {\cal M}_{f}\;{\cal M}_{g}\;\cdots$, where each element is represented by a symplectic map ${\cal M}_{f}$, with $(\exp\mbox{:f$_{2}$:})$ describing paraxial behavior, $(\exp\mbox{:f$_{3}$:})$ and $(\exp\mbox{:f$_{4}$:})$ describing second- and third-order aberrations, etc. Lastly, the product of maps for two successive elements can be factorized using the BCH theorem (\ref{eq:BCH}). Consequently, a map {\it action} ${\cal M}: z^{in} \rightarrow z^{out}$ can be explicitly constructed (by drawing tremendous advantage from symbolic manipulation algorithms).

Applications of Lie-transform methods to the development of electron microscopes and linear colliders, for example, can 
lead to the cancellation of spherical aberration (always present in axially symmetric electromagnetic systems) by the introduction of a sextupole beam-line correcting element. This cancellation can be explicitly calculated by Lie-transform methods and a consequent spot-size reduction by at least one order of magnitude can be observed (e.g., Dragt presented a case where the spot-size was reduced from about 3 \AA $\;$to less than 0.1 \AA). Similar applications in accelerator physics, for example, provide accelerator designers with the capability of following charged particles through more than 
one millions of turns around a storage ring.

\vspace*{0.2in}

\no
{\sf IV. LAGRANGIAN METHODS}

\vspace*{0.2in}

Lagrangian (or variational) formulations of fluid and plasma physics provide powerful guiding principles for the derivation of exact and reduced dynamical equations and the construction of their associated invariants and conservation laws. The present Section summarizes the talks presented on applications of Lagrangian methods in fluid and plasma physics.

Before proceeding with our summary, we point out that near-identity transformations also exist in fluid dynamics (e.g., ideal magnetohydrodynamics). Here, near-identity transformations are expressed in terms of the Lie operator $T_{\epsilon}$ defined as
\begin{equation} 
T_{\epsilon}\;\chi({\bf x},t) \;=\; \exp(\epsilon\;\vb{\xi}\bdot\nabla)\;\chi({\bf x},t) \;=\; \chi({\bf x} + \epsilon\,\vb{\xi}),
\label{eq:fluid}
\end{equation}
where $\chi({\bf x},t)$ is an arbitrary fluid scalar field and the vector field $\vb{\xi}$ represents the ideal fluid displacement from a {\it reference} (or unperturbed) position ${\bf x}$ to the {\it exact} (or perturbed) position ${\bf x} + \epsilon\,\vb{\xi}({\bf x},t)$ of a fluid element. Since near-identity transformations are invertible, we find the inverse operator $T_{\epsilon}^{-1} = \exp(-\,\epsilon\,\vb{\xi}
\bdot\nabla)$. For Hamiltonian fluid dynamics (with unperturbed fluid velocity ${\bf u}_{0}$), the unperturbed evolution operator is $d_{0}/dt = \partial/\partial t + {\bf u}_{0}\bdot\nabla$ while the perturbed evolution operator is $d_{\epsilon}/dt = \partial/\partial t + 
{\bf u}_{\epsilon}\bdot\nabla$, where the perturbed fluid velocity can be written as 
\begin{equation}
{\bf u}_{\epsilon}({\bf x},t;\, \vb{\xi})\bdot\nabla \chi({\bf x},t) \;=\; T_{\epsilon}^{-1} 
\left( \frac{d_{0}}{dt}\;T_{\epsilon}\;\chi({\bf x}, t) \right).
\label{eq:delta_u}
\end{equation}
Additional expressions for perturbed fluid quantities, which can also be derived by 
Lie-transform methods, include the perturbed fluid density 
\begin{equation}
\rho_{\epsilon}({\bf x},t;\,\vb{\xi}) \;=\; {\sf T}_{\epsilon}^{-1}\rho_{0}({\bf x},t)\;\;{\rm det}\left[\nabla \left(
{\sf T}_{\epsilon}^{-1}{\bf x}\right)\right], 
\label{eq:delta_rho}
\end{equation}
the perturbed fluid entropy 
\begin{equation}
s_{\epsilon}({\bf x},t;\,\vb{\xi}) \;=\; {\sf T}_{\epsilon}^{-1}s_{0}({\bf x},t), 
\label{eq:delta_s}
\end{equation}
and the perturbed (magnetic) vector potential 
\begin{equation}
{\bf A}_{\epsilon}({\bf x},t;\,\vb{\xi}) \;=\; {\sf T}_{\epsilon}^{-1}{\bf A}_{0}({\bf x},t) \;-\; \nabla\left(
{\sf T}_{\epsilon}^{-1}{\bf A}_{0}({\bf x},t)\right)\bdot\delta_{\epsilon}{\bf x}, 
\label{eq:delta_A}
\end{equation}
where $\delta_{\epsilon}{\bf x} \equiv {\sf T}_{\epsilon}^{-1}{\bf x} - {\bf x}$ and we have omitted the gauge term $\nabla({\sf T}_{\epsilon}^{-1}{\bf A}_{0}\bdot\delta_{\epsilon}
{\bf x})$ in Eq.~(\ref{eq:delta_A}). The expressions (\ref{eq:delta_rho})-(\ref{eq:delta_A}) can also be derived from the Lie derivative ${\cal L}_{\vb{\xi}}$ of appropriate differential forms, e.g., $\rho_{\epsilon}\,
d^{3}x = \exp(-\epsilon\,{\cal L}_{\vb{\xi}})[\rho_{0}\,d^{3}x]$. Note that the constrained variations of the fluid and electromagnetic fields $\delta\psi^{a} = (\delta\rho, \delta{\bf u}, \delta s, \delta{\bf A})$ used in the standard variational formulation of ideal magnetohydrodynamics \cite{EP_review,Newcomb} are defined from $\psi_{\epsilon}^{a} = (\rho_{\epsilon}, {\bf u}_{\epsilon}, s_{\epsilon}, {\bf A}_{\epsilon})$, as given in Eqs.~(\ref{eq:delta_u})-(\ref{eq:delta_A}), as $\delta\psi^{a}(\vb{\xi}) \equiv (d\psi_{\epsilon}^{a}/d\epsilon)
|_{\epsilon = 0}$. For example, $(d{\bf u}_{\epsilon}/d\epsilon)_{\epsilon = 0} = (\partial_{t} + {\bf u}_{0}\bdot\nabla)\vb{\xi} - \vb{\xi}\bdot\nabla{\bf u}_{0} = \delta{\bf u}$ yields the well-known expression for the constrained (Eulerian) variation of the fluid velocity ${\bf u}$ \cite{Newcomb}.

Darryl Holm presented a variational principle for Euler-Poincar\'{e} equations of geophysical fluid dynamics amenable to asymptotic analysis based on the procedure of Lagrangian averaging \cite{GLM}. Holm pointed out that the challenge of geophysical fluid modeling is to model large-scale (or slow) dynamics while retaining the effects of small-scale (or fast) dynamics, as represented by fluid turbulence. This situation is reminiscent of the single-particle case where fast time scales (e.g., orbital time scales associated with gyromotion or bounce motion of a charged particle in a strong magnetic field) are asymptotically eliminated by Lie-transform methods \cite{RGL}. In the present case, two averaging procedures (Eulerian and Lagrangian) offer opposing characteristics. Here, the Lagrangian average of a fluid quantity 
$\chi({\bf x},t)$, denoted as $\ov{\chi}({\bf x},t)$ and taken at constant Lagrangian coordinate ${\bf x}_{0}$, is defined as $\ov{\chi}({\bf x},t) = \langle \chi({\bf x} + \epsilon\,\vb{\xi},\; t)\rangle$, where $\langle\;\rangle$ denotes an Eulerian average (performed at a fixed spatial location ${\bf x}$) and ${\bf x} + \epsilon\,\vb{\xi}({\bf x},t)$ represents the exact position of a fluid element whose mean position is ${\bf x}$ (we note that our notation is slightly different than that of Ref.~\cite{GLM}). Hence, Eulerian averaging (taken at a fixed spatial location) commutes with the spatial gradient and partial time derivative and, consequently, the momentum-conservation form of hydrodynamical equations is invariant under Eulerian averaging. Lagrangian averaging (or {\it path}-averaging taken by following a fluid element) commutes with the advective time derivative moving with the fluid and, consequently, Lagrangian averaging preserves important conservation laws (e.g., Kelvin circulation and potential vorticity). Unfortunately, Eulerian averaging does not preserve these conservation laws while Lagrangian averaging does not commute with spatial gradients and is (by definition) history dependent. Nevertheless, a generalized Lagrangian mean (GLM) theory for Euler-Poincar\'{e} equations can be developed within the context of constrained variational principles \cite{GLM}. 

The description of large-scale (slow) dynamics while retaining the effects of the small-scale (fast) dynamics is provided by a wave, mean flow interaction (WMFI) model, constructed by asymptotic analysis based on two dimensionless parameters $(\alpha,\epsilon)$ associated with wave amplitude ($\alpha$) and eikonal representation $(\epsilon)$ of the wave fields. For this purpose, we replace $\epsilon\vb{\xi}({\bf x},t)$ in Eqs.~(\ref{eq:fluid})-(\ref{eq:delta_s}) with $\alpha\,\wt{\vb{\xi}}(\epsilon{\bf x},\epsilon t)\;\exp[i\epsilon^{-1}\theta(\epsilon{\bf x},\epsilon t)] + {\rm c.c.}$ so that the perturbed fluid fields $\rho_{\epsilon}$, ${\bf u}_{\epsilon}$, and $s_{\epsilon}$ are replaced with $\rho_{(\alpha,\epsilon)}$, ${\bf u}_{(\alpha,\epsilon)}$, and $s_{(\alpha,\epsilon)}$, respectively. These expressions are then inserted into the variational principle for geophysical fluid dynamics and expanded in powers of $\alpha$ and $\epsilon$. The mean-flow Lagrangian is obtained at zeroth order in $\alpha$ and $\epsilon$, while at order $\alpha^{2}$, terms representing ray-optics, self-modulation (dispersion); WKB stability can be collected at zeroth order, first order, and second order in $\epsilon$, respectively. Some approximations of the WMFI variational principle include \cite{GLM}: (i) the derivation of the mean-flow dynamics by setting $\alpha = 0$ and varying with respect to the mean fields; (ii) variation with respect to the wave fields only yields equations suitable for linearized spectral analysis (with unperturbed mean-flow dynamics included); and (iii) variation with respect to mean fields and wave fields, after phase averaging is performed, yields the WMFI dynamics.

Lazar Friedland presented his work on an averaged variational principle for multi-phase nonlinear waves excited and controlled by synchronized nonlinear mode conversion by slow (adiabatic) passage through resonance \cite{phase_locking}. 
The analysis focused on small periodic perturbations of periodic solutions of the Sine-Gordon (SG) equation \cite{SG} $u_{tt}(x,t) - u_{xx}(x,t) + \sin u(x,t) = \epsilon\,f(x,t)$. By using an eikonal representation for the perturbation, Friedland showed that, when the amplitude of the driving perturbation is above a certain threshold, one can excite a single phase nonlinear waveform synchronized with the drive. Furthermore, as the driving parameters vary in time and/or space, the emerging phase-locked waveform slowly evolves near the solution space of the unperturbed problem and, if desired, becomes highly nonlinear. The oscillatory solutions of the nonlinear SG equation are the plasma and breather oscillations and Friedland showed how each nontrivial nonlinear solution of the SG equation created from zero and controlled by a small forcing. Lastly, Friedland presented an outline of the theory and numerical simulation results of this excitation process based on multi-phase Whitham's averaged variational principle.

Eliezer Hameiri presented a review of the variational formulation of the Energy-Casimir stability analysis of plasma equilibria with general flows \cite{Hameiri}. Beginning with a single-fluid model (Hall-MHD), Hameiri identified two stream functions $\psi$ and $\varphi$ associated with axisymmetric plasma equilibrium. The first stream function $\psi$ is the familiar magnetic flux function defined from the magnetic equation ${\bf B}\bdot\nabla\psi = 0$ while the second stream function $\varphi$ defines flow surfaces through the relation 
${\bf u}\bdot\nabla\varphi = 0$. We note that a general plasma equilibrium with flows requires that its entropy $s$ satisfy ${\bf u}\bdot\nabla s = 0$, i.e., the entropy $s(\varphi)$ is a function of the flow stream function $\varphi$. Moreover, Hameiri showed that constant-entropy surfaces are also defined from an application of a modified version of Ertel's theorem, which leads to the condition ${\bf B}^{*}\bdot\nabla s = 0$, where ${\bf B}^{*} = {\bf B} + a^{-1}
\,\nabla\btimes{\bf u}$ is related to the curl of the canonical momentum of the single-fluid (with $a = e_{i}/m_{i}$). Hence, in an axisymmetric plasma equilibrium with flows, we may write ${\bf B} = \nabla\zeta\btimes\nabla\psi + I\,\nabla\zeta$ and ${\bf B}^{*} = \nabla\zeta\btimes\nabla\varphi + I^{*}\,\nabla\zeta$, where $I$ and $I^{*}$ are functions of $\psi$ and $\varphi$ ({\it our notation}). After identifying all the constants of the motion for a given plasma equilibrium with flows, a variational principle is developed \cite{Steinhauer}, which yields a description of plasma equilibrium with flows in terms of two coupled Grad-Shafranov equations for $\psi$ and $\varphi$. Hameiri also discussed extensions of this variational formulation of equilibria with flows to the multi-species magnetofluid case and the classical fluid case.

Jean-Luc Thiffeault presented work on a unified variational formulation of the nonlinear stability analysis of static \cite{static} and stationary \cite{stationary} plasma equilibria. In particular, the methods of {\it Eulerianized} Lagrangian Displacements (ELD) \cite{Newcomb} [see Eqs.~(\ref{eq:delta_u})-(\ref{eq:delta_A})], Dynamical Accessibility (DA) \cite{DA}, and Energy-Casimir (EC) \cite{EC} are shown to be equivalent in assessing the stability of magnetohydrodynamic (MHD) equilibria. Thiffeault showed how sufficient conditions for the stability of MHD equilibria can be obtained by using the Dynamical Accessibility method \cite{Morrison_RMP}. The use of dynamical accessibility ensures that the physical perturbations preserve the natural constraints of the system imposed by the Hamiltonian structure. Here, the first and second variations of the fluid fields $\zeta^{\alpha}$ are expressed as $\delta\zeta^{\alpha} = \{ {\cal G},\; \zeta^{\alpha}\}$ and $\delta^{2}\zeta^{\alpha} = \frac{1}{2}\,\{ {\cal G},\; \{ {\cal G},\; \zeta^{\alpha}\}\;\}$, where ${\cal G} = \int \zeta^{\alpha}\,\chi_{\alpha}\;d^{3}x$ is a generating functional and $\{\;,\;\}$ is an appropriate Poisson bracket structure. From the {\it ansatz} $\vb{\dot{\xi}} = \rho\;\nabla
\chi_{1} - \chi_{2}\,\nabla s + {\bf B}\btimes\nabla\btimes\vb{\chi}_{3}$, Thiffeault showed that the ELD and DA methods 
are equivalent if the functions $\chi_{1}$, $\chi_{2}$, and $\vb{\chi}_{3}$ span the space of $\vb{\dot{\xi}}$ (and {\it vice versa}). Lastly, Thiffeault commented on the fact that this procedure is a generalization of the work of Newcomb \cite{Newcomb} and Arnold \cite{Arnold}, among others, who make use of Lagrangian displacements.

\vspace*{0.2in}

\no
{\sf V. LAGRANGIAN CHAOS}

\vspace*{0.2in}

Even in the presence of dissipation, a knowledge of Hamiltonian dynamics improves our understanding of the role played by dissipative dynamics. John Finn presented numerical results of work on self-consistent Lagrangian chaos (i.e., chaotic advection) in locking bifurcations 
in two-dimensional shear flows. Here, two-dimensional shear flows are modeled by the velocity field ${\bf v} = \nabla\phi\btimes\wh{z}$ and the vorticity scalar field $\omega = \wh{z}
\bdot\nabla\btimes{\bf v} = -\,\nabla^{2}\phi$ in terms of the scalar field $\phi(x,y,t)$. 
Two-dimensional fluid motion $\dot{{\bf x}} = {\bf v}$ is thus described as a Hamiltonian system $\dot{x} = \partial\phi/\partial y$ and $\dot{y} = -\,\partial\phi/\partial x$, with $\phi$ acting as the Hamiltonian. As is well known, the fluid motion is integrable if the Hamiltonian is time independent, or can be made time independent by a change 
of variable, e.g. a change of frame; if the Hamiltonian is time dependent, Lagrangian chaos is possible. Lagrangian chaos 
is defined by the property of fluid elements following chaotic trajectories while Eulerian chaos is defined in terms of turbulence at a fixed location; as a result of Lagrangian chaos, if a passive scalar field is injected into a chaotic flow, the passive scalar field is advected, stretched, and folded by the chaotic nature of the flow \cite{LC_flow}. Finn pointed out that Lagrangian chaos does not imply Eulerian chaos (although it enhances transport) and that even laminar periodic flows can lead to complex Lagrangian trajectories with statistical properties similar to turbulent flows.

Next, Finn investigated the self-consistent Lagrangian chaos in a fluid system consisting of a nonlinear wave due to a shear-flow (Kelvin-Helmholtz) instability and an externally imposed sinusoidal perturbation; the fluid was taken to be 
two-dimensional and incompressible. Self-consistent Lagrangian chaos was observed in the sense that the vorticity was {\it actively} advected with the flow \cite{footnote_2}. The Navier-Stokes system describing the fluid was studied by direct numerical simulation (involving symplectic integration techniques). In the region of parameter space studied (moderately large Reynolds number), Lagrangian chaos in the presence of smooth flows is relevant, and in this range the bifurcations which occur are low dimensional. The main features observed during direct numerical simulations are locking and unlocking bifurcations, the latter related to the self-consistent Lagrangian chaos \cite{LC_1}. Lastly, Finn discussed the relation with self-consistent Lagrangian chaos in shear flows supporting two Kelvin-Helmholtz modes of very different phase velocities \cite{LC_2} (where locking is impossible).

\vspace*{0.2in}

\no
{\sf VI. SUMMARY}

\vspace*{0.2in}

The Mini-Conference was successful in bringing together experts from a variety of fields in physics who actively apply Lagrangian and Hamiltonian methods in their research. In his talk, Morrison highlighted the surprisingly prevalent lexicon of finite-dimensional Hamiltonian systems (that include such terms as integrability, Poincar\'{e} sections, KAM invariant tori, intrinsic stochasticity, and Chirikov's overlap criterion). Indeed, such concepts are now part of our language, independently of whether we use Lagrangian or Hamiltonian methods in our research. Kaufman also reminded us of the story of the discovery of the third (drift) adiabatic invariant by Northrop. According to Kaufman, Edward Teller called in the Sherwood theory group, and posed the problem of how only 2 (known) invariants were insufficient to explain the confinement of individual particles in the recently discovered van Allen belt. As Kaufman recalled, Ted Northrop on his own deduced the drift invariant, and then co-authored a paper with Teller \cite{Northrop}.

The Mini-Conference also included a poster session during which many presentations were made on applications of Hamiltonian and Lagrangian methods in fluid and plasma physics. Presenters from France, Sweden, Germany, the United Kingdom, Russia, and the United States provided ample evidence of the vitality of this field of research in fluid dynamics, laser-plasma interactions, space plasma physics, and theoretical fluid and plasma physics.

\vspace*{0.2in}

\no
{\sf ACKNOWLEDGMENTS}

The Authors wish to thank all presenters for their participation at the Mini-Conference. The occasion of the 
Mini-Conference allowed the participants to celebrate Allan Kaufman's 75$^{th}$ birthday. On behalf of all the 
participants, the co-organizers wish to acknowledge Allan's leadership and contributions in plasma physics over nearly 
fifty years and to thank him for the inspiration he has provided to many of us.

The Mini-Conference was supported by the Division of Plasma Physics of the American Physical Society.


\begin{thebibliography}{99}
\bibitem{footnote_1} Note that a Casimir invariant $C(\vb{\zeta})$ can be added to the Hamiltonian $H(\vb{\zeta},t)$ 
without changing the Hamilton's equations $d\zeta^{k}/dt = \{ \zeta^{k},\; H\}$ since $\{\zeta^{k},\; (H + C)\} = \{\zeta^{k},\; H\}$.
\end{thebibliography}
\end{document}